\RequirePackage{ifpdf}
\ifpdf 
\documentclass[pdftex]{sigma}
\else
\documentclass{sigma}
\fi

\begin{document}

\renewcommand{\PaperNumber}{001}

\FirstPageHeading

\ShortArticleName{The Differential Form Method for Finding Symmetries}

\ArticleName{The Differential Form Method for Finding Symmetries}

\Author{B. Kent HARRISON}

\AuthorNameForHeading{B.K. Harrison}

\Address{Physics and Astronomy Department, Brigham Young
University, Provo, Utah 84602, USA}
\Email{\href{mailto:bkentharrison@comcast.net}{bkentharrison@comcast.net},
\href{mailto:bkh@byu.edu}{bkh@byu.edu}}

\ArticleDates{Received July 20, 2005; Published online August 03, 2005}

\Abstract{This article reviews the use of differential forms and
Lie derivatives to find symmetries of differential equations, as
originally presented in Harrison and  Estabrook
1971~\cite{harrison&estabrook}.  An outline of the method is
given, followed by examples and references to recent papers using
the method.}


\Classification{22E70; 34A26; 35A25; 35A30; 53Z05}

\section{Introduction}

In 1969--70, Frank Estabrook and the present author found a method
for finding symmetries of differential equations using
differential forms and Cartan's formulation of differential
equations~\cite{harrison&estabrook}.  (This will be called Paper~I.)
 It was not something we were searching for; rather, we were
simply trying to understand how the symmetries of Maxwell's
equations could be found from the differential form version of
those equations.  Once we realized that the key to symmetries was
the use of the Lie derivative, it became clear how to apply this
to all differential equations.  An outline of the method, with
examples, will be given here.  A few computer prog\-rams that use it
will be mentioned, along with a number of published papers on
symmetries which have used it.

The reader may wonder at the order of the names of the authors on
our original paper.  Since we had done roughly equal amounts of
work on the research, the order was determined by the flip of a
coin.

The method proceeds as follows.  We consider a set of partial
differential equations, defined on a differentiable manifold $M$
of $n$ independent variables and $m$ dependent variables.
(Ordinary differential equations constitute a special case; we
will mention those later.)  We define the partial derivatives of
the dependent variables as new variables (prolongation) in
sufficient number to write the equations as a set of first order
equations, thus extending the manifold to a manifold~$M'$.  Then
we can formulate those as differential forms.  We speak of the set
of forms, representing the equations, as an ideal $I$.  It is to
be closed.

We can recapture the original set of equations by two operations:
we specialize to a submani\-fold by letting the dependent variables
be functions of the independent variables (``sectioning'') and then
we set the pullback of the forms to zero (``annulling'').  The
resultant equations are the original set of first order partial
differential equations, and the submanifold is a solution
submanifold.  (All calculations are local and we do not use any
manifold structure except diffe\-ren\-tiability.)

Lie derivatives of geometrical objects, like tensors, are
associated with symmetries of those objects.  The Lie derivative
of a geometrical object carries it along a path, determined by
a~vector~$\mathbf{v}$, in its manifold.  If the Lie derivative
vanishes, then the vector $\mathbf{v}$ represents the direction of
an infinitesimal symmetry transformation in the manifold.   A
differential form is a~type of tensor (totally antisymmetric on
the indices of the components), so it has a Lie derivative.  We
may construct the Lie derivative (symbolized by $\pounds$) of the
forms in the ideal $I$.  Setting the Lie derivative of these forms
equal to zero should therefore represent symmetries~--- except for
one thing.  When we make an infinitesimal transformation away from
the original variables, we require that the new form of the
differential equations should vanish~--- but the old form must also
vanish.  Thus we want the original forms in $I$ to vanish, but
also their Lie derivative must vanish when that happens.  In other
words, we require that the Lie derivatives of the forms in $I$ to
be linear combinations of those forms themselves~--- and when they
vanish, then the Lie derivatives also vanish.  We can express this
by writing $\pounds_{\mathbf{v} } I = 0 \pmod{I} $, or
\begin{gather} \pounds_{\mathbf{v} } I  \subset I. \label{eq1}
\end{gather}
This is satisfied by letting the Lie derivative of each
differential form in $I$ be a linear combination of the forms in
$I$.

Equation \eqref{eq1} will contain a number of Lagrange
multipliers, the coefficients of the forms in their linear
combinations.  Those are to be eliminated.  Once they are
eliminated, there remains a set of linear homogeneous first order
equations for the components of $\mathbf{v}$ in $M'$, which are
the symmetry generators.  The equations are simply the determining
equations for the symmetries of the original set of differential
equations, considered as point transformations in $M'$.

We note, in this set, that the derivatives of the components of
$\mathbf{v}$ (the generators) will be taken with respect to both
dependent variables (including prolonged ones) and independent
variables.  One often assumes that the generators for the
independent variables (often denoted by~$\xi$ and~$\eta$) are
functions only of the independent variables (there are
exceptions.)  The determining equations will usually show that
feature promptly.

Some examples of familiar equations will be presented to show how
the method works.  Some of this material was presented by the
author at the second Kiev symmetry conference in 1997 and can be
found in its Proceedings \cite{harrison}.  This will be denoted as
paper II.

In paper I, we adopted the term ``isovector'' for the vector
$\mathbf{v}$, describing a symmetry transformation in $M'$, even
though the term had been used elsewhere in the physics literature.
We did not think there would be any confusion.  A number of
authors thus refer to this method as the ``isovector'' method.

\section{Lie derivatives of differential forms}

First we note some simple features of Lie derivatives of
differential forms.  (See paper I.)
\begin{enumerate}
\itemsep=0pt
\item[(1)] Lie differentiation
preserves the rank of a form.
\item[(2)] The Lie derivative of a
coordinate is simply the component of $\mathbf{v}$ in that
direction:
\[
\pounds_{\mathbf{v} } x^i  = v^i .
\]
\item[(3)] The Lie derivative of a function on $M'$ ($0$-form) is simply
its directional derivative:
\[
\pounds_{\mathbf{v} } f = \mathbf{v} (f) = v^i  f_{,i}.
\]
(Commas represent partial derivatives.  Sometimes they will
omitted when the context is clear.)
\item[(4)] The Lie derivative of a
wedge product obeys the Leibniz rule (the subscript $\mathbf{v}$
may be suppressed where it is not necessary):
\[
\pounds (\alpha \wedge \beta) = (\pounds \alpha) \wedge \beta +
\alpha \wedge (\pounds  \beta).
\]
\item[(5)] The exterior derivative $d$ and the Lie derivative $\pounds$
commute.  In particular,
\[
\pounds_{\mathbf{v} } dx^i = d (\pounds_{\mathbf{v} } x^i ) = d
v^i ;
\]
the Lie derivative of the differential of a variable equals the
differential of the corresponding component of $\mathbf{v}$.
\end{enumerate}

\section{The one-dimensional heat equation}

We write the one dimensional heat equation
\begin{gather}
u_{xx} = u_t    \label{eq2}
\end{gather}
as a first order set of equations by defining a new variable $w$:
\begin{gather}
u_x = w , \qquad w_x  = u_t .   \label{eq3}
\end{gather}
The variables are $x$, $t$, $u$, and $w$.  We construct two $2$-forms
by inspection:
\begin{gather}
\alpha = du \wedge dt - w dx \wedge dt,  \nonumber\\
\beta = dw \wedge dt + du \wedge dx  \label{eq4}
\end{gather}
($\alpha$ is a contact form.)  If we ``section'' these
forms --- specialize to a submanifold $u = u(x,t)$ and $w =
w(x,t)$ --- we get
\[
\alpha = (u_x  dx + u_t dt) \wedge  dt - w dx \wedge dt = (u_x -
w) dx \wedge dt
\]
and
\[
\beta = (w_x dx + w_t dt) \wedge dt + (u_x dx + u_t dt) \wedge dx
= (w_x - u_t) dx \wedge dt
\]
where we have used the antisymmetry of $1$-forms.  We now ``annul''
these forms --- set them equal to zero --- obtaining Eqs.~\eqref{eq3},
the original first-order set of equations.  The forms $\alpha$ and
$\beta$ in Eqs.~\eqref{eq4} now constitute the ideal $I$ of forms
representing the heat equation~\eqref{eq2}.

We note that $I$ is not unique; we may as well represent the heat
equation by defining $z = u_t$ and constructing an ideal $I'$ with
a $1$-form
\[
\gamma = -du + w dx + z dt,
\]
its exterior derivative
\[
d\gamma = dw \wedge dx + dz \wedge dt,
\]
and the $2$-form
\[
\delta = dw \wedge dt - z dx \wedge dt.
\]
We note that $\alpha = - \gamma \wedge dt$ and $\beta = \delta -
\gamma \wedge dx.$

We work first in the ideal $I$.  Write the Lie derivatives of
$\alpha$ and $\beta$ as linear combinations of themselves.  Expand
the Lie derivatives by the rules above.  We also drop the wedge
product $\wedge$ and the subscript $\mathbf{v}$ on $\pounds$ to
save writing.
\begin{gather}
\pounds \alpha = \pounds (du \, dt - w dx \, dt) \nonumber \\
\phantom{\pounds \alpha}{} = (\pounds du) dt + du (\pounds dt) - (\pounds w) dx \, dt  - w (\pounds dx) dt - w dx (\pounds dt) \nonumber \\
\phantom{\pounds \alpha}{} = dv^u  dt + du \, dv^t - v^w dx \, dt - w dv^x dt - w dx \, dv^t \nonumber \\
\phantom{\pounds \alpha}{} = \lambda_1 (du \, dt - w dx \, dt) + \lambda_2 (dw \, dt + du \, dx). \nonumber
\end{gather}
The $\lambda_i$ are $0$-forms (functions).  Expand the $dv^i$ by
the usual chain rule, since the $v^i$ are functions in $M'$, using
all four variables.  Since $dt \, dt = 0$, etc., by the
antisymmetry of $1$-forms, some terms drop out.   We have
\begin{gather}
(v^u_{,u} du + v^u_{,x} dx + v^u_{,w} dw) \,
dt + du \, (v^t_{,t} dt + v^t_{,x} dx + v^t_{,w} dw) \nonumber \\
\qquad {}- v^w dx \, dt - w (v^x_{,x} dx + v^x_{,u} du + v^x_{,w} dw) \, dt \nonumber \\
\qquad {}- w dx \, (v^t_{,t} dt + v^t_{,u} du + v^t_{,w} dw)  \nonumber \\
\qquad {}= \lambda_1 (du \, dt - w dx \, dt) + \lambda_2 (dw \, dt
+ du \, dx). \nonumber
\end{gather}
There will be $4!/2! 2! = 6$ basis $2$-forms ($dx \, dt$, $dx \, du$,
$dx \, dw$, $dt \, du$, $dt \, dw$, and $du \, dw$.)  We equate the
coefficients of these $2$-forms to get
\begin{gather}
v^u_{,x}  - v^w  - w(v^x_{,x} + v^t_{,t}) = -w \lambda_1, \nonumber \\
- v^t_{,x}  - w v^t_{,u} = - \lambda_2,  \nonumber \\
- w v^t_{,w}  = 0, \nonumber \\
- v^u_{,u} - v^t_{,t} + w v^x_{,u} = - \lambda_1,  \nonumber \\
 - v^u_{,w} + w v^x_{,w} = - \lambda_2,  \nonumber \\
 v^t_{,w} = 0. \nonumber
\end{gather}
Eliminating the Lagrange multipliers $\lambda_i$ gives us one half
of the determining equations:
\begin{gather}
v^t_{,w} = 0,     \nonumber \\
v^t_{,x} + w v^t_{,u} = v^u_{,w} - w v^x_{,w},   \nonumber \\
v^u_{,x} - v^w - w v^x_{,x} = - w v^u_{,u} + w^2 v^x_{,u}.
\nonumber
\end{gather}
Expansion of $\pounds \beta$ gives us the other half.  One quickly
sees from them that $v^t$  is a function of $t$ only and that
$v^x$ is a function only of $x$ and $t$.  Further calculation
gives the usual six generators plus addition of an arbitrary
solution.  Exponentiation of the transformation proceeds by
setting $\mathbf{v} \cdot \gamma = 0$ (contraction of $\mathbf{v}$
and $\gamma$, symbolized by a dot) and solving, where $w$ and $z$
are replaced by their values as derivatives of $u$ (the usual
method).

Another way to proceed, which removes the need for the
multipliers, is to use $\alpha = 0$ to replace $du \, dt$,
anywhere that combination occurs in the expansion of the Lie
derivatives of $\alpha$ and~$\beta$, by $w dx \, dt$, and to use
$\beta = 0$ to replace $dw \, dt$ by $- du \, dx$.  This may save
considerable work in complicated cases, especially in cases where
not all forms in the ideal are of the same rank.  In those cases,
some of the Lagrange multipliers may need to be forms (of rank
greater than zero) themselves in order for the right hand sides to
be of the same rank as the left hand sides, and that means that
there may be very many coefficients to be eliminated.  If one can
avoid that, labor may be saved.

One can also use the ideal $I'$ for the heat equation.  (This is
the technique used in paper I.)  There are now five variables in
$M'$:  $x$, $t$, $u$, $w$, and $z$. In $I'$, there is only one $1$-form
$\gamma$, and so its Lie derivative equation is simple:
\begin{gather}
\pounds \gamma = \lambda \gamma,   \label{eq5}
\end{gather}
where $\lambda$ is a multiplier.  One can expand the Lie
derivative by an identity for any form $\omega$, using the
contraction operator:
\begin{gather}
\pounds_{\mathbf{v} } \omega  = d(\mathbf{v} \cdot \omega) +
\mathbf{v} \cdot d\omega.  \label{eq6}
\end{gather}
Write $F = \mathbf{v} \cdot \gamma$, which is a function, and
expand Eq. \eqref{eq5} using Eq. \eqref{eq6} (with $\omega =
\gamma$ ) and identities for contraction (e.g., $\mathbf{v} \cdot
(dx \, dy) = v^x dy - v^y dx $) (see paper I).  We get
\begin{gather}
F = -v^u  + w v^x + z v^t   \label{eq7}
\end{gather}
and
\begin{gather}
\pounds_{\mathbf{v} } \gamma = dF + v^w  dx - v^x dw + v^z dt - v^t dz \nonumber \\
\phantom{\pounds_{\mathbf{v} } \gamma}{}= \lambda \gamma = \lambda (-du + w dx + z dt).
 \nonumber
\end{gather}
Expand $dF$ with the chain rule, equate coefficients, eliminate
$\lambda$ and use Eq.~\eqref{eq7}, and we get all generators $v^i$
in terms of $F$ and its derivatives (subscripts on $F$ are
derivatives):
\begin{gather}
v^x = F_w , \qquad  v^t = F_z , \qquad v^u = - F + w F_w + z F_z , \nonumber \\
v^w = - F_x - w F_u , \qquad  v^z  =  - F_t - z F_u . \nonumber
\end{gather}
The exterior derivative of Eq.~\eqref{eq5} is
\[
d (\pounds \gamma) = \pounds (d\gamma) = d\lambda \wedge \gamma +
\lambda d\gamma
\]
so that the Lie derivative of $d\gamma$ is also in the ideal.
There is now only one equation left:
\[
\pounds \delta = \lambda_1 \delta + \lambda_2 d\gamma + \tau
\wedge \gamma,
\]
where the $\lambda_i$ are $0$-forms and $\tau$ is an arbitrary
$1$-form with five terms.  The term in $du$ in $\tau$ can be
eliminated by substituting from $\gamma$, and that drops out. This
procedure also gives the standard determining equations.

\section{Computer programs}

There are a few computer programs which use this technique.  Some
of these were written, in REDUCE, by D.G.B. Edelen~\cite{edelen1}.
The programs are probably still available from
Lehigh University.  Other programs were written by Gragert,
Kersten, and Martini, also in REDUCE.  They published several
works which developed and used this software, including a program
for symbolic integration of overdetermined systems \cite{gragert,
kersten&gragert1,gragertetal,kersten1,kersten2,gragert&kersten}.
Problems treated in references
\cite{kersten&gragert2,kersten&martini,kersten3} are studies of a
nonlinear diffusion equation
\[
\triangle (u^{p + 1}) + k u^q  =  u_t,
\]
(where $\triangle$ is the Laplacian), the massive Thirring model,
and the Federbush model.  The present author used one of the
programs with E.D. Fackerell to explore a relativity problem, in
unpublished work, and one of Fackerell's students, Ben Langton,
used it for his Ph.D. dissertation on certain solutions of the
Einstein equations~\cite{langton}.

Another useful computer program is \textbf{liesymm}, a program
found in MAPLE, based on a paper by Carminati et al.~\cite{carminati}.
It works quite well, and there is an additional
program called \textbf{autosimp} in MAPLE which does some
integration of the determining equations, although the integration
may not be complete.  These programs are discussed briefly in
Refs.~\cite{heck,char}. A student of the author's, David Neilsen,
did a master's thesis with~\textbf{liesymm} on Einstein's
equations~\cite{neilsen}.

An extensive review of symbolic software was done by Hereman
\cite{hereman} in 1997.  A nice table of programs is provided.  No
specific distinction is made in that paper between the traditional
method and the differential form method.

\section{Nonlinear Boltzmann equation}

This example is actually listed as an example in \textbf{liesymm}
in Maple 9.5, although it is not worked out.  The equation is:
\[
u_{xt} + u_x + u^2 = 0.
\]
Possible ideals are $I'$ with five variables, $x$, $t$, $u$, $p = u_x$,
$q = u_t$:
\begin{gather}
\alpha = du - p dx - q dt, \nonumber \\
d\alpha = - dp \, dx - dq \, dt, \nonumber \\
\beta = - dt \, dq + (p + u^2) dx \, dt \nonumber
\end{gather}
or $I$ with four variables, $x$, $t$, $u$, $p$:
\begin{gather}
\gamma = du \, dt - p dx \, dt, \nonumber \\
\delta = dp \, dx + (p + u^2 ) dt \, dx. \nonumber
\end{gather}
The calculation is quite similar to that for the heat equation.
There are four generators.

\section{Vacuum Maxwell equations}

From paper I we write the usual $3$-forms that represent the
vacuum Maxwell equations in rectangular coordinates.  Subscripts
represent components.
\begin{gather}
\alpha = dE_x \, dx \, dt + dE_y \, dy \, dt + dE_z \,
dz \, dt + dB_x \, dy \, dz + dB_y \, dz \, dx + dB_z \, dx \, dy,   \nonumber \\
\beta = dB_x \, dx \, dt + dB_y \, dy \, dt + dB_z \, dz \, dt -
dE_x \, dy \, dz - dE_y \,  dz \, dt - dE_z \, dx \, dy. \nonumber
\end{gather}
We simplify these forms by defining $\gamma = \alpha + i \beta$
and $(A, B, C) =$ (cyclic $E_k + i B_k) = \mathbf{h}$.  Then
(paper I) we can write
\[
\gamma = d\mathbf{h} \cdot (d\mathbf{r} \, dt - (1/2) i
d\mathbf{r} \times d\mathbf{r})
\]
or
\begin{gather}
\gamma = dA \, (dx \, dt - i dy \, dz) + dB \, (dy \, dt - i dz \,
dx) + dC \, (dz \, dt - i dx \, dy). \label{eq8}
\end{gather}
The forms in the ideal will then be $\gamma$ and $\gamma * $,
where the star represents complex conjugate.  The variables will
be $t$, $x$, $y$, $z$, $A$, $B$, $C$, $A^{\ast}$, $B^{\ast}$, $C^{\ast} $. The
generators for the coordinates $t$, $x$, $y$, $z$ will be real.  The
equations for the Lie derivatives are then
\[
\pounds \gamma = \lambda \gamma + \mu \gamma *
\]
and its complex conjugate.

In paper I the determining equations were worked out by using a
vector-dyadic formalism.  Here we use Eq.~\eqref{eq8} for
$\gamma$, which is a little clearer.  We work with $3$-forms in
ten variables, so that there are $10! / 7! 3! = 120$ different
basis $3$-forms.  There are two equations, for $\pounds \gamma$
and $\pounds \gamma * $, so that we apparently have 240 equations.
However, 120 of them are simply the complex conjugates of the
others.  So we just look at the $\pounds \gamma$ equation.  We see
immediately by inspection that 3-forms with all terms being
$d$(field variable) do not appear.  Terms of the form $d$(field)
$\wedge$ $d$(field*) $\wedge$ $d$(coordinate) yield only equations
for the derivatives of the coordinate generators with respect to
the complex conjugate fields (which are zero).  Thus the
coordinate generators do not depend on the complex conjugate
fields, nor (by complex conjugation) on the fields.  Thus they
depend only on the coordinates themselves. It is also easy to show
that the field generators depend only on the fields and not on
their complex conjugates.

This reduces the number of equations to 22--18 with one field
$1$-form and two coordinate forms and four with three coordinate
forms.  From the first set we get the conformal Killing equations
plus some expressions for the derivatives of the field generators.
From the second we get equations for the field generators that are
the Maxwell equations themselves.  Solution of the determining
equations gives the 17-generator set given in paper I (15
conformal Killing vectors, a scale change on the fields, and a
duality change on the fields, plus the addition of an arbitrary
solution).  Steeb \cite{steeb} presents other symmetries besides
these, which depend on the derivatives of the fields.  (Steeb and
collaborators also treat various versions of the Dirac equation
\cite{steebetal}.)

\section{Nonlinear Poisson equation}

We consider the equation:
\[
u_{xx} + u_{yy} + u_{zz} = f(u),
\]
where $f(u)$ is an undetermined function.  Subscripts represent
derivatives.  We define $r = u_x$, $s = u_y$, $t = u_z$.
Then
\[
r_x + s_y + t_z = f(u).
\]
The ideal $I$ consists of these forms:
\begin{gather}
\alpha = - du + r dx + s dy + t dz,  \nonumber \\
d\alpha = dr \, dx + ds \, dy + dt \, dz,    \nonumber \\
\beta = dr \, dy \, dz + ds \, dz \, dx + dt \, dx \, dy - f(u) \,
dx \, dy \, dz. \nonumber
\end{gather}
There are seven variables.

We may approach this problem by defining a function $H =
\mathbf{v} \cdot \alpha$, as we did with the heat equation.  The
Lie derivative of $\alpha$ gives all the generators in terms of
$H$ and its derivatives, as before.  Then the only equation we
have left is that for $\pounds \beta$, a $3$-form.  But equating
it to a linear combination of $\alpha$, $d\alpha$, and $\beta$ is
messy.  The multiplier of $d\alpha$, a $2$-form, must itself be a
$1$-form, which will have six coefficients (we do not include a
term in $du$, because that can be replaced by~$\alpha$, and
$d\alpha \wedge \alpha$ can be included in the $\alpha$ term.) The
multiplier of $\alpha$ must be a $2$-form~--- and again we can
eliminate $du$ terms because they can be replaced by $\alpha$, and
$\alpha \wedge \alpha = 0$.  But that still leaves 15
coefficients.  The multiplier of $\beta$ will be a single
coefficient.  That totals 22 coefficients that must be eliminated.

So we consider an easier way.  We define a new ideal $I'$, made up
of four $3$-forms:  $\beta$, $\alpha \, dy \, dz = (-du + r dx)
\, dy \, dz, \, \alpha \, dz \, dx$, and $\alpha \, dx \, dy$.
The latter three forms are equivalent to $\alpha$ alone.  The Lie
derivative of each $3$-form must be a linear combination of all
four, thus giving four multipliers in each equation to be
eliminated.  The equations are much simpler; it is easy to
eliminate four multiplier coefficients in each equation than 22,
even though there are now still $4 \times 4 = 16$~multipliers.  The
easiest procedure is to write out the equation for the Lie
derivative of $\alpha \, dy \, dz$, eliminate the multipliers to
get a set of determining equations and then to permute $x$, $y$, $z$
(and $s$, $t$) cyclically.  One quickly gets the result that the
generators for $x$, $y$, $z$, and $u$ are functions only of $x$, $y$, $z$,
$u$, and the generators for $r$, $s$, and $t$ are given in terms of a
function which is precisely the $H$ defined above, $H = \mathbf{v}
\cdot \alpha  = - v^u + r v^x  + s v^y + t v^z$.

The equation for $\pounds \beta$ now has four multipliers, which
are easily eliminated.  We find quickly that $v^x$, $v^y$, $v^z$
depend only on $x$, $y$ and $z$ and  that they obey the Killing
equations.  The generators for $r$, $s$, and $t$ are written out
easily, and one ends up with a single equation involving $f(u)$
and $f'(u)$.  Solution of that equation for the given $f$ then
leads to the final result.

There is a small technical point.  We did not include $d\alpha$ in
$I'$, even though we did represent $\alpha$ as three $3$-forms.
Should we have done so?  The answer is no.  If we take the
exterior derivative of the $3$-forms $\alpha \, dy \, dz$, etc. we
get terms like $dr \, dx \, dy \, dz$ --- in other words, just
$d\alpha \, dy \, dz$.  We get three of those equations, which are
equivalent to the equation for $\pounds d\alpha$.  The determining
equations for $I$ and $I'$ give the same result.

A similar treatment is used by Satir \cite{satir}, who writes a
set of two-dimensional bosonic membrane equations as eight
$3$-forms.  He then uses a REDUCE program together with the EXCALC
differential geometry package to find a 12 parameter group.  He
remarks that the use of diffe\-ren\-tial forms enabled the calculation
to go more quickly that the conventional method.

\section{Nonlinear diffusion equation}

This an equation treated in paper II, originally due to
Fushchych~--- a nonlinear diffusion equation with an additional
condition.  We can write a $1$-form as was done above, its
exterior derivative, and a $4$-form for the main field equation.
In paper II, it was assumed \textit{a priori} that the generators
for the coordinates depend only on the coordinates and that those
for the derivatives of the field were linear in those derivatives.
It then turns out that much of the analysis of the Lie derivative
of the $4$-form can be done by inspection.

\section{One-dimensional compressible fluid dynamics}

The equations considered here are (see paper I):
\begin{gather}
\rho_t  + (\rho u)_x = 0,  \nonumber \\
\rho u_t + \rho u u_x + c^2 \rho_x = 0, \nonumber
\end{gather}
where isentropic flow is considered so that the pressure is only a
function of the density, $P = P(\rho)$, and $c^2 = dP/d\rho$.  The
generators for $x$ and $t$ include a $\rho$- and $u$-dependent
case, which turns out to give the hodograph transformation.

\section{Nonclassical symmetries}

One can generalize the ideal $I$ by including contractions of
$\mathbf{v}$ with some of the differential forms.  An example of
this was provided in paper I for the heat equation, in which it
was shown that one gets the equations for ``nonclassical
symmetries'' of that equation, the same equations found by Bluman
and Cole in 1969 \cite{bluman}.  While this technique has not been
explored in detail by this author, Webb has studied this set of
equations --- referred to as a coupled nonlinear Burgers-heat
equations system --- with differential forms and has searched for
B\"{a}cklund transformations for the set \cite{webb1}.

\section{Ordinary differential equations}

We consider an example:
\[
y'' = f(x, y, y'),
\]
where the prime indicates differentiation with respect to $x$.  We
put $z = y'$ and write two $1$-forms, $\alpha = dy - z dx$ and
$\beta = dz - f(x,y,z) dx$.  There are three variables.  The Lie
derivative equation for $\alpha$ is
\[
\pounds \alpha = d v^y - v^z dx - z dv^x = \lambda_1 (dy - z dx) +
\lambda_2 (dz - f dx).
\]
There are three equations, for the coefficients of $dx$, $dy$, and
$dz$; elimination of the multipliers gives a single equation,
which is an expression for $v^z$.  The Lie derivative equation for
$\beta$ gives another single equation.  We assume that the
generators $v^x$ and $v^y$ (usually written as $\xi$ and~$\eta$,
respectively) are functions only of $x$ and $y$.  In that case,
$v^z$  becomes the usual extended generator for $z = y'$ and the
remaining equation is the usual determining equation for $\xi$ and
$\eta$, as given, e.g., in Stephani \cite{stephani}.

\section{Advantages of using differential forms}

These have been treated in paper II, but are reviewed here. The
method is easy to apply.  One simply writes all equations as first
order equations; the differential forms can be written by
inspection.  Calculations may be long because of the necessity of
introducing the Lagrange multipliers; however, one can choose the
ideal to minimize this.  One can make use of symmetries of form
(e.g., cyclic symmetry of coordinates), or one can use the forms
to substitute for certain terms in the Lie derivative expansion,
thus removing the need for multipliers.
Independent variable generators may easily be
considered as functions of the independent variables only (just assume that
and that simplifies the expansion of the differentials of those generators).

\section{Other examples}

We mention here some research papers in which the differential
form method is used.  Papachristou generalized the method to
vector-valued or Lie algebra-valued differential forms and treated
the two-dimensional Dirac equation and the Yang--Mills free-field
equations in Minkowski spacetime \cite{pap1} (as part of a Ph.D.
dissertation with the author.)  This was later used to investigate
self-dual Yang--Mills equations, which work showed connections
between symmetry and integrability (in the form of B\"{a}cklund
transformations) of those equations \cite{pap2,pap3,pap4}. Waller,
in three similar papers, treats nonlinear diffusion equations (or
reaction-diffusion equations) arising in plasma physics
\cite{waller1,waller2,waller3}.  He uses the technique of writing
a 1-form and contracting it with~$\mathbf{v}$, as done in the
second treatment of the heat equation above.

Edelen has developed the theory of the differential form method
extensively.  His computer programs have already been mentioned.
At least two books \cite{edelen2,edelen3} and several papers
\cite{edelen4, edelen5, edelen6,edelen7} explore the use in
differential forms in physics, including the method discussed
here.  In papers \cite{edelen5,edelen6} he considers a method of
characteristics in any number of dimensions, using isovector
treatments. With this he can write parametric solutions of
differential equations. One equation he considers is
\cite{edelen6}
\[
u_t \, u_x = 4u.
\]
He gives a solution for the equation as an initial value problem:
if $u(x,0) = \alpha(x)$, with $\alpha '(x) \neq  0$, then
\begin{gather}
x = z + \alpha (z) (\exp (4\tau) - 1)/ \alpha '(z),  \nonumber \\
t = (1/4) \alpha '(z) (\exp (4\tau) - 1),   \nonumber \\
u = \alpha (z) \exp (8\tau), \nonumber
\end{gather}
where $z$ is a parameter and $\tau$ is an arbitrary function.
Instead, if one defines $r = u_x$, one can write a simple $1$-form
for the equation:
\[
 \alpha = -du + r dx + (4u/r) dt.
\]
Then $\pounds \alpha = \lambda \alpha$ gives equations which yield
most of the generators in terms of a function $F = \mathbf{v}
\cdot \alpha$, which satisfies the linear first order equation
(subscripts are derivatives)
\[
F_t + 4 F_r  + (4u/r^2) F_x + (8u/r) F_u = 4 F/r.
\]
The special solution $F = u/r$ seems to give Edelen's solution,
although not all details are worked out yet and it is a little
uncertain.  In Ref.~\cite{edelen7} he considers ``inverse''
isovector methods.

Webb et al consider nonlinear Schr\"odinger equations for a type of
MHD waves, using the differential form method \cite{webbetal}.  He
also analyzes a nonlinear magnetic potential equation, with
conservation laws, with the Liouville equation as a special case
\cite{webb2}.  Pakdemirli and others treat boundary layer
equations for non-Newtonian fluids, including arbitrary shear
stress, power law fluids, and other models \cite{pak1,pak2}.
\c{S}uhubi and others, in a number of papers, consider general
approaches to equations of balance and other equations
\cite{suh1,suh2,suh3,pak3,suh4,ozer,suh5}.  A number of these
discuss equivalence groups, as a generalization of symmetry
groups.  One paper with Ozer \cite{ozer} treats nonvacuum Maxwell
equations with nonlinear constitutive relations.  Another
discusses steady boundary layer flow past a semi-infinite flat
plate \cite{suh3}.

Bhutani and Bhattacharya study $n$-dimensional Klein--Gordon and
Liouville equations with an interesting approach \cite{bhu1}.
Various types of diffusion equations are treated in Refs.~\cite{bhu2,bhu3,chow1}.
Viscoelastic-viscoplastic rods are
studied in Ref.~\cite{chow2} and power law creep in Ref.
\cite{delph}.  Equations of meteorology, here meaning steady
two-dimensional incompressible inviscid flow with a Coriolis term,
are studied in Ref.~\cite{vij}.   Hu considers the principal
chiral model \cite{hu}, using differential forms and ideas from
Ref.~\cite{pap3}.  An interesting paper is that by Barco, who
shows for a second-order hyperbolic or parabolic differential
equation, with one dependent variable and two independent
variables, that an isovector can be used to generate a similarity
solution by using a particular Cauchy characteristic vector field
\cite{barco}.

Nonlinear thermoelasticity was treated by Kalpakides \cite{kal}.
His work is related to that of \c{S}uhubi
\cite{suh1,suh2,suh3,pak3,suh4,ozer,suh5}.  Harnad and Winternitz
considered a generalized nonlinear Schr\"odinger equation,
\[
i z_t  + z_{xx} = f(z, z* ),
\]
with attention to both symmetries and B\"{a}cklund transformations
\cite{harnad}.

\subsection*{Acknowledgements}

Thanks to Anatoly Nikitin for the suggestion to present this topic
and to Vyacheslav Boyko for help with the literature.

\LastPageEnding

\end{document}